\documentclass[12pt,a4paper]{article}
\pdfoutput=1  
\usepackage{amssymb}
\usepackage{amsmath}
\usepackage{amsfonts}
\usepackage{enumerate}
\usepackage{float}
\usepackage{natbib}
\usepackage{verbatim}
\usepackage{graphicx}
\usepackage{lscape}
\usepackage[autostyle]{csquotes}
\usepackage{bbm}
\usepackage[dvipsnames]{xcolor}
\usepackage{pdflscape,array,booktabs}
\usepackage[a4paper,bindingoffset=0.2in,left=1in,right=1in,top=1in,bottom=1in,footskip=.5in]{geometry}
\usepackage{caption}
\usepackage{siunitx,booktabs} 
\usepackage{hyperref}

\restylefloat{table}

\graphicspath{ {images/} }   

\DeclareMathAlphabet{\pazocal}{OMS}{zplm}{m}{n}

\begin{document}

\title{Robust Ranking of Happiness Outcomes: A Median Regression Perspective 
\thanks{An earlier version of this paper had been circulated under the title \textquotedblleft Have Econometric Analyses of Happiness Data Been Futile? A
Simple Truth About Happiness Scales\textquotedblright. We thank Valentina
Corradi, Arthur Lewbel, Tong Li, Oliver Linton, Andrew Oswald, Jo\~{a}o
Santos Silva, Matt Shum and seminar participants at Academia Sinica,
European Meeting of the Econometric Society (2019), and the London School of
Economics for helpful comments and discussions.}  
\thanks{{\fontsize{2}{4}Contacts: Le-Yu Chen: Institute of Economics, Academia Sinica, No. 128, Section 2, Academia Road, Nankang,
Taipei, 115, Taiwan, lychen@econ.sinica.edu.tw; Ekaterina Oparina: Centre for Economic Performance, London
School of Economics, Houghton Street, London, WC2A 2AE, UK, e.oparina@lse.ac.uk; Nattavudh Powdthavee: Warwick
Business School, University of Warwick, Coventry, CV4 7AL, UK, nattavudh.powdthavee@wbs.ac.uk; Sorawoot
Srisuma: National University of Singapore,
Department of Economics 1 Arts Link AS2 06-02, Singapore, 117570, s.srisuma@nus.edu.sg.}}}
\author{\quad\quad\quad Le-Yu Chen \quad\quad\quad \\
\quad\quad\quad Academia Sinica \quad\quad\quad \and \quad\quad\quad
Ekaterina Oparina \quad\quad\quad \\
\quad\quad\quad London School of Economics \quad\quad\quad \and \quad
Nattavudh Powdthavee \quad \\
\quad Warwick Business School \quad \and \quad\quad\quad Sorawoot Srisuma
\quad\quad\quad \\
\quad\quad\quad NUS and University of Surrey\quad\quad\quad }
\date{June 6, 2022 }
\maketitle

\begin{abstract}
Ordered probit and logit models have been frequently used to estimate the mean ranking of happiness outcomes (and other ordinal data) across groups. However, it has been recently highlighted that such ranking may not be identified in most happiness applications. We suggest researchers focus on median comparison instead of the mean. This is because the median rank can be identified even if the mean rank is not. Furthermore, median ranks in probit and logit models can be readily estimated using standard statistical softwares. The median ranking, as well as ranking for other quantiles, can also be estimated semiparametrically and we provide a new constrained mixed integer optimization procedure for implementation. We apply it to estimate a happiness equation using General Social Survey data of the US.


\end{abstract}

\vspace{1.5in}

\setcounter{page}{1}\newpage

\section{Introduction}

The study of human happiness has been cited as one of the fastest growing
research fields in economics over the last two decades (\cite%
{kahneman_developments_2006, clark_relative_2008, stutzer_recent_2013}). By
looking at what socioeconomic and other factors predict (or cause) people to
report higher or lower scores on a subjective well-being (SWB) scale,
researchers have been able to add new insights to what have become standard
views in economics. For example, studies of job and life satisfaction have
shown that people tend to care far more about relative income rather than
absolute income (\cite%
{clark_satisfaction_1996,ferrer-i-carbonell_income_2005}), while
unemployment is likely to hurt less when there is more of it around (\cite%
{clark_unemployment_2003,powdthavee_are_2007}). The use of SWB data has
therefore enabled economists to test many of the assumptions in conventional
economic models that might have been untestable before the availability of
proxy utility data (e.g., \cite{di_tella_preferences_2001,
stevenson_subjective_2013, gruber_cigarette_2006, boyce_money_2013}). It has
also led many policy makers to start redefining what it means to be
successful as a community and as a nation (\cite{kahneman_toward_2004,
stiglitz_measurement_2009, de_neve_sdgs_2020}).

One objective of happiness research is to understand the determinants of SWB
and how they compare across groups. The predominant approach to the SWB data
analysis is either through linear regression (OLS) or ordinal parametric
methods (e.g. ordered probit or logit); see \cite%
{ferrericarbonell_how_2004} for a comprehensive discussion on the validity of both approaches. Conclusions are then drawn, as is common in
applied fields of economics and other social sciences, based on conditional
or unconditional mean comparisons using these estimates. Recent studies,
however, have suggested that there might be a serious methodological problem associated with these standard estimation approaches.

The issue traces to the fact that SWB is an ordinal measure. For example,
consider the 3-point ordinal happiness scale in the General Social Survey
(GSS). In the GSS, respondents were asked whether they were
\textquotedblleft 1. not too happy\textquotedblright , \textquotedblleft 2.
pretty happy\textquotedblright , or \textquotedblleft 3. very
happy\textquotedblright . We know that a score of 3 is higher than a score
of 2 or 1. But we cannot interpret the extent of which each category is
higher than another without further assumptions. Suppose we are interested
in comparing a particular statistic between two ordinal variables such as
ranking the mean SWB of two different groups. The rank order of any
statistic is identified if that order relation remains unchanged when the
ordinal variables undergo any increasing transformation. In particular, this
implies, the mean ranking of \textit{ordinal} variables is identified if and
only if there holds between them a first order stochastic dominance (FOSD)
relation. However, a FOSD relation between two variables does not always
exist in which case the mean ranking between them is not identified. In
contrast, the mean ranking between \emph{cardinal} variables is always
identified as long as they have finite first moment.

When an ordinal variable is used as the dependent variable in a linear
regression, the sign of a slope parameter may be used to indicate the mean
ranking across groups. However, the mean ranking is not identified if the
sign of such parameter can change by relabelling the ordinal outcomes that
preserve the order of the categories. See \cite{schroder_revisiting_2017}
for a detailed analysis on this issue. A somewhat analogous problem exists
in the context of an ordinal regression model. To see
this, as we shall do throughout this paper, we interpret observed ordinal
outcomes through a threshold crossing latent variable model. In this
setting, Bond and Lang (2019, BL hereafter) point out in Section II.B of their paper
that any pair of
latent variables that follow a continuous two-parameter distribution and
have different means, there is a FOSD relation between them if and only if
their variances are identical. This result\footnote{
See Theorem 1 in \cite{bond_2014} for a general statement of this result.
}, in particular, implies the mean rankings in ordered probit and logit
models are identified if and only if the respective latent 
variables are homoskedastic. 
The following example illustrates its
relevance in a simple context of male-female happiness comparison using a
probit model.

\bigskip

\textsc{Example 1: }Suppose we observe $Y$ and $D$ that respectively denote
reported happiness from a 3-point scale and a female gender dummy so that 
\begin{equation*}
Y=\left\{ 
\begin{array}{c}
1 \\ 
2 \\ 
3%
\end{array}%
\begin{array}{c}
\text{if }H\leq 0 \\ 
\text{if }0< H\leq 1 \\ 
\text{if }H>1%
\end{array}%
\right. \mathbf{,}\text{ for }H|D\sim N\left( \beta _{0}+\beta _{1}D,\sigma
^{2}\left( D\right) \right) .
\end{equation*}
If $0<\Pr \left[
D=1\right] <1$, then we can identify $\beta
_{1}=E\left[ H|D=1\right] -E\left[ H|D=0\right] $. Let $\beta _{1}\neq 0$ and suppose we want to use the sign of $\beta _{1}$ to determine group comparison. Consider these two
situations.

\begin{itemize}
\item \textit{Homoskedastic case} $\left( \sigma ^{2}\left( 1\right) =\sigma
^{2}\left( 0\right) \right) $. Then the mean ranking between $H|D=1$\ and $H|D=0$
is identified by the sign of $\beta _{1}$. I.e., if $\beta _{1}\left.
>\left( <\right) \right. 0$ then $E\left[ \tau \left( H\right) |D=1\right] -E%
\left[ \tau \left( H\right) |D=0\right] \left. >\left( <\right) \right. 0$
for all increasing function $\tau $.$\ $

\item \textit{Heteroskedastic case} $\left( \sigma ^{2}\left( 1\right) \neq
\sigma ^{2}\left( 0\right) \right) $. Then the mean ranking between $H|D=1$\ and $%
H|D=0$ is not identified. In this case $\beta _{1}$ is not informative for
the mean ranking. I.e., if $\beta _{1}\left. >\left( <\right) \right. 0$
then there must exist an increasing function $\tau $ such that $E\left[ \tau
\left( H\right) |D=1\right] -E\left[ \tau \left( H\right) |D=0\right] \left.
<\left( >\right) \right. 0$.
\end{itemize}

\bigskip 

\noindent BL demonstrates the practical implication of non-identification by
taking on nine of the most well-known findings from the happiness
literature. They test and reject the homoskedasticity hypothesis for 8 out
of 9 cases\footnote{%
\cite{bond_sad_2019} did not have individual level data used in \cite{ludwig_neighborhood_2012} for comparing happiness between the Control and Experimental
groups in the Moving to Opportunity program. They used shares of different
happiness levels reported in the paper to back out the mean and variance of
latent happiness under normality assumption, from which they can perform a
transformation to reverse the result that suggests subjects in
the experimental group were happier than those in the control group.}, and
then they show conclusions previously drawn from mean comparisons can be
reversed by applying some exponential transformation to the latent happiness
or SWB variable in all cases.

The discussion above raises an important question about how we should interpret
empirical results obtained from ordered probit and logit models. This
concern is highly relevant because many applications use standardized ordered probit and logit models
, which assume the conditional variance of latent happiness are respectively $1$ and $\frac{\pi ^{2}}{3}$, when the homoskedasticity assumption may not hold in the data. By contrast, the mean ranking in any generalized ordered probit or logit model 
that presumes heteroskedasticity cannot be identified.  Nevertheless, irrespective of whether the mean ranking is identified, many existing empirical findings appear
intuitive and have been widely accepted in the literature. This suggests we
may be able to learn something about group ranking even if the mean ranking
is fundamentally not identified.

Our main goal is to provide a pragmatic view on how to interpret results
estimated from ordered logits and probits regardless whether or not the mean
rank is identified, as well as to suggest an alternative method to compare ordinal
variables generally. Our argument focuses on using the median as a statistic
for comparisons instead of the mean. We have the following messages.

\begin{enumerate}
\item The median ranking of ordinal variables is identified under weak
conditions without requiring FOSD.

\item The median ranking is identified in probit and logit models. It can in
fact be identified by the conditional means of latent variables of these
models. The economic implications of prior results based on the mean ranking
are therefore robust when interpreted as the median rank even when FOSD does
not hold.
\end{enumerate}

\noindent We focus on the median, an alternative well-known measure of central
tendency, because it is \textquotedblleft equivariant\textquotedblright\ to
all increasing transformations. I.e., letting $Med(Z)$ denote the median of
a random variable $Z$ and $\tau $ be an increasing function, then $\tau
\left( Med(Z)\right) =Med(\tau \left( Z\right) )$. Therefore once a median rank for any pair of latent variables has been established for a particular cardinalization, it cannot be reversed by any monotone transformation.

\bigskip

\textsc{Example 1 (cont}\textit{'}\textsc{): }Let $Med\left( H|D\right) $
denote the conditional median of $H$ given $D$. By symmetry of the normal
distribution, $Med\left( H|D\right) =E\left[ H|D\right] =\beta _{0}+\beta
_{1}D$. Therefore $\beta _{1}$ identifies $Med\left( H|D=1\right) -Med\left(
H|D=0\right)$. Furthermore, by equivariance, the sign of $\beta _{1}$ is the
same as that of $Med\left( \tau \left( H\right) |D=1\right) -Med\left(
\tau \left( H\right) |D=0\right) $ for every increasing function $\tau $. Thus
the median rank between men and women is identified by the sign of $\beta
_{1}$.

\bigskip

The continuation of Example 1 highlights perhaps the most empirically important point
of our paper. That is: the median rank of an ordinal variable in an ordered
probit or logit model can be identified even when the mean rank is not;
furthermore, the median rank is identified by the sign of a model parameter
that is routinely estimated in practice. This is due to the fact that the normal and logistic distributions are symmetric and the median identified in logit and probit models coincides with the mean. Specifically, researchers can practically
perform group comparisons in the same fashion as previously\footnote{For examples, with Stata, \texttt{oprobit} and \texttt{ologit} can be used to estimate standardized probit and logit models respectively, and \texttt{oglm} can be used to estimate their generalized counterparts where users can specify the form of heteroskedasticity.}, the only change is
to interpret rankings in terms of the median instead of the mean. To see why
this simple change of stance can be very powerful, let us consider again the
empirical illustrations in BL. There, BL show all of their ordered probit
estimates that allow for heteroskedasticity deliver qualitatively the same
conclusions as in previous studies under homoskedasticity before they apply
exponential transformations to reverse the mean rankings. Therefore, under
the median interpretation, their estimates would support existing results in
the happiness literature rather than dispute them.

Our paper also explores robust estimation of the median. 
When the analysis is conducted with probit or logit models, it is implicitly assumed that latent happiness follows normal or logistic distribution. Furthermore, in the heteroskedastic case, when the conditional variance of happiness distribution is allowed to vary across respondents with different characteristics, the researcher typically chooses a functional form of the conditional variance. If these distributional or functional form assumptions do not hold, fully parametric models are misspecified and subsequent estimators would be inconsistent. From the econometrics literature, \cite{lee_median_1992}
has shown it is possible to identify and consistently estimate the median
semiparametrically without any parametric distributional assumption as well
as functional form of heteroskedasticity. Estimating Lee's estimator in practice, however, can be a very challenging task. His estimator, in particular, is a generalization of the maximum
score estimator (MSE), which \cite{manski_semiparametric_1985} introduced for estimating a binary choice model and is well-known for difficult implementation, for estimating a model with multiple categories. 

Another contribution in this paper is that we provide a new and
computationally efficient procedure to compute Lee's estimator. Our computational approach is based on the method of mixed integer optimization, which was used by \cite{florios_exact_2008} to implement Manski's MSE. We adapt their procedure designed for a binary choice to a multiple choice setting. Importantly, our procedure can also be used to estimate at any quantile in addition to the median. Since every quantile is equivariant to increasing transformations, quantile ranks are identified under weak conditions. We illustrate in the paper how results estimated at other quantiles can provide additional insights to supplement the conventional mean-median analysis. We apply our estimation procedure and show that a standard happiness equation structure continues to hold under our new approach (\cite%
{blanchflower_well-being_2004}).

While there are compelling statistical reasons for pursuing median ranks
over mean ranks, especially when the former is identified and the latter is
not, a question of normative interest is whether, instead of the mean, a policy maker would prefer to
use the median or other quantiles to represent aggregate well-being measures. A recent article by \citet{sechel_share_2021} makes an
interesting argument that the emphasis on using averages, which implicitly indicates a preference for the conventional mean welfare maximization, may not be a
sustainable social goal with finite resources. She provides anecdotal
evidence in support of a sufficientarian welfarism view where a policy maker's aim is to have a \textit{sufficient} level of welfare reached instead of improving welfare for everyone. More specifically, she suggests a planner may want to analyze a measure constructed from
averaging \textquotedblleft headcounts\textquotedblright\ of reported
well-being scores that is no less than a targeted threshold. I.e., the
threshold corresponds to the $\alpha$-quantile of the reported well-being
score where $\alpha $ is the average headcount in that sample. The ordinal
regression framework we study in this paper can be particularly useful for a
sufficientarian planner as it is able to perform group comparisons within a population at any targeted quantile level. In
addition, from the decision theory literature, an economic agent whose goal
is to maximize the quantile of their utility distribution has a well-established
foundation. We refer readers to \citet{manski1988} and, for more recent
developments, to \citet{rostek2010} and \citet{de2019} for decision
theoretical justifications.

We emphasize that our work is not a critique of BL. Their theoretical point
that identification of the mean ranking of ordinal data in familiar
parametric models is possible only when homoskedasticity holds is insightful
and cannot be disputed. We hope, however, to prevent the potentially extreme
interpretation of their results that nothing about ordinal ranking can be
learnt from popular parametric models and the value of prior works that used
probit or logit models rests on the knife-edge condition whether the model is homoskedastic or not.

BL also point out another challenge for ordinal data analysis that is
relevant for median ranking. They question the often assumed notion in
empirical studies of a common reporting function that puts the latent
variables from different groups on the same cardinal scale. In a threshold
crossing model, this corresponds to the possibility that threshold values
are heterogeneous across groups. We also maintain the common threshold
assumption in the present paper. One may relax this assumption by using a parametric
compound hierarchical ordered response model (see \cite{KinMurSal04}). To the best of our knowledge, extensions to semiparametric ordinal median regression models with heterogeneous thresholds have not yet been developed and would be an important topic for further research.

The remainder of this paper proceeds as follows. Section 2 gives an account on how statistical analysis for discrete ordinal data have been developed in economics and other disciplines, and on some recent developments on estimating the median. Section 3 introduces an ordered response model and formalizes our argument to identify the median and other quantile ranks. Section 4 compares parametric and semiparametric estimation, and introduces the mixed integer optimization approach to the computation of the median and quantile estimators in a semiparametric ordered response model. Section 5 revisits estimation of a happiness equation using GSS data. Section 6 concludes. The Appendix provides the details of the mixed integer optimization based implementation algorithm in the context of semiparametric median and quantile estimation problems. 

\section{Discrete ordinal data analysis: a brief review}

We consider discrete ordinal outcome that represents an individual's ordered
categorical response in the data. 
The defining property of an ordinal variable is that there is a \textit{rank
order} over values it can take but the distances between these values are
arbitrary and carry no information. A discrete ordinal variable can
therefore, in constrast to cardinal variables, be put on a scale like $%
\left\{ 1,2,..,J\right\} $ without any loss of generality. Such data
measurements are common in social and biomedical sciences. Examples include
individual happiness (unhappy, neither happy nor unhappy, happy), severity
of injury in the accident (fatal injury, incapacitating injury,
non-incapacitating, possible injury, and non-injury), and lethality of an
insecticide (unaffected, slightly affected, morbid, dead insects) among many
others.

Ordinal data analysis in a regression framework is widely acknowledged to
have been co-founded by two independent sources. 
One originates from the contribution of \cite{mckelvey_statistical_1975},
who developed the well-known ordered probit model to study Congressional
voting on the 1965 Medicare Bill. 
The other is due to \cite{mccullagh_regression_1980}, who focused on
modelling proportional odds and proportional hazards that become prominent
in the biomedical fields. Huge literature on ordinal data analysis has since
grown from these influential works. We refer interested readers to \cite%
{greene_hensher_2010} and reference therein for the developments in social
science, \cite{agresti_modelling_1999} for the medical science, and \cite%
{ananth_regression_1997} for epidemiology.

Researchers from different fields take different approaches to analyzing
ordinal data. Applied researchers in biomedical fields pay a great deal of
attention to choosing an appropriate model for their data (goodness of fit)
but place less importance on the interpretation of individual parameters
(e.g. coefficients in a generalized linear model). On the other hand,
researchers in social sciences often focus on the model parameters.
Economists, for example, typically work with linear regression or ordered
probit and logit models that are very convenient for interpreting
parameters, especially in a setting with many covariates. One research area in economics that has utilized ordered discrete response models the most is perhaps the economics of well-being. Given the ordinal nature of SWB data, many early and classic studies in this field have exclusively used ordered probit and logit models to analyze different predictors of human happiness, including unemployment (\citet{clark_unhappiness_1994}), political institutions (\citet{frey_happiness_2000}), income inequality (\citet{alesina_inequality_2004}), and relative income (\citet{ferrer-i-carbonell_income_2005}). More recently, many researchers have also been using linear regression models to conduct their analysis. For example, OLS model has been used to estimate the relationship between SWB and macroeconomic factors such as inflation and unemployment rate (\citet{di_tella_preferences_2001}), comparison income (\citet{luttmer_neighbors_2005}), and fertility (\citet{kohler_population_2005}). One particular advantage for using linear models is the ease in incorporating and accommodating fixed effects when panel data are used.

The linear regression method treats ordinal variables as if they were
cardinal. Several studies have shown linear regression and ordered
probit/logit models could deliver similar qualitative results empirically.
For example, in a study of vehicle driver injury severity, \cite%
{gebers_exploratory_1998} compared both OLS and ordered logit estimation
results and found that estimated coefficients of both models were of the
same sign and generally agreed in magnitude and statistical significance. In an empirical analysis of the effect of economics sanctions, \cite%
{major_timing_2012} compared the OLS and ordered probit estimates and found
that the results were similar across these two estimating models. See also 
\cite{ferrericarbonell_how_2004}, who found both the OLS and ordered probit
models produced comparable results in their study on the sources of
individual well-being.

One practical issue with the linear regression approach is the dependency on
the reported scale. Particularly, it is possible for OLS estimates to change signs if the reported data are monotonically transformed. See, e.g. \cite{schroder_revisiting_2017} for a sufficient condition of this; also see \citet{kaiser_how_nodate} for a sufficient condition for which the signs of OLS estimates cannot be reversed. Some recent works have suggested complementing OLS estimates obtained from the reported scale with those that undergo some transformations as a sensitivity analysis. E.g., \citet{bloem_how_nodate} proposes a class of one-parameter monotone functions and suggesting to estimate a set of OLS estimates indexed by that parameter; \citet{bloem_analysis_2021} suggest dichotomizing the dependent variable, e.g. into high and low happiness groups using the median of reported happiness to form the split, to mitigate the dependency of the reported scale. 

For latent variable threshold crossing ordinal regression models, the
regression coefficients are by construction invariant to any order
preserving relabeling of the ordinal outcome. These parameters naturally
have an interpretation of the conditional mean difference of latent
variables across different groups. Their signs can thus be used to identify the
group mean ranking whenever the ranking order is invariant for all
increasing transformations of the latent variable. This requirement amounts
to the condition of there being a FOSD relation between latent variable
distributions of the groups. In a fully parametric model such as ordered probit and logit, given a specific form of heteroskedasticity and assuming the means across groups differ, a hypothesis test of homoskedasticity would determine if there is a FOSD relation. It is also possible to test the null against a nonparametric alternative in this setting because the conditional variance function can be nonparametrically identified within a probit or logit framework (e.g. see \citet[][Lemma 1]{oparina_analyzing_2021} ). In these cases, rejecting homoskedasticity means there is no FOSD relation. This knife-edge condition makes identification of the mean rank order in probit and logit models very fragile as BL have illustrated.
We refer readers to their paper for further discussions. More generally one
can test FOSD directly in a semiparametric or nonparametric setting that does not assume distributional assumption of the latent variable. See
e.g. \cite{carneiro_estimating_2003}, \cite{cunha_identification_2007}, \cite%
{lewbel_constructing_1997, lewbel_semiparametric_2000}, \cite%
{lewbel_simple_2007}, \cite{honore_semiparametric_2002} and \cite%
{KaplanZhuo2021b}.

In this paper we propose that a natural alternative for ranking ordinal
outcomes is to focus on the median instead of the mean. For commonly used
ordered response models such as the ordered probit and logit, the median and
mean are identical. Maximum likelihood estimation of the median in these
models can therefore be performed as readily as the mean using standard
statistical softwares. The median can also be estimated semiparametrically
without distributional assumption. The seminal work of \cite%
{manski_maximum_1975, manski_semiparametric_1985} develops the maximum score
estimation in this setting for a binary choice model and \cite%
{lee_median_1992} extends this to multiple ordered choice data. While
theoretically appealing, performing maximum score estimation for the
discrete choice model is computationally difficult. More specifically, the
ordered response median regression estimator proposed by \cite%
{lee_median_1992} is a solution to a non-smooth and non-covex least absolute
deviation (LAD) optimization problem. The corresponding LAD objective
function, which is akin to Manski's maximum score objective function, is non-convex -- it is 
piecewise constant with numerous local solutions. The computational
challenges for solving the maximum score estimation problem is well noted in
the econometrics literature (e.g. see \cite{manski_operational_1986}, \cite%
{pinkse_computation_1993}, \cite{skouras_algorithm_2003}).

Recently \cite{florios_exact_2008} propose a mixed integer optimization
(MIO) based approach for the computation of maximum score estimators. In
particular, they show that Manski's binary choice maximum score estimation
problem can be equivalently reformulated as a mixed integer linear
programming problem (MILP). \cite{chen_best_2018} provide an alternative
MILP formulation that complements the approach of \cite{florios_exact_2008}
for solving the maximum score estimation problem. These reformulations
enable exact computation of maximum score estimators through modern
efficient MIO solvers. Well-known numerical solvers such as CPLEX and Gurobi
can be used to effectively solve the MIO problems. We refer readers to \cite%
{bertsimas_optimization_2005} and \cite{conforti_integer_2014} for recent
and comprehensive texts on the MIO methodology and applications.

The estimators of \cite{manski_semiparametric_1985} and \cite%
{lee_median_1992} have been shown to be consistent under very weak
conditions. On the other hand, maximum score type estimators converge at a
cube-root rate and have non-standard asymptotic distributions. See \cite%
{kim_cube_1990} and \cite{seo_local_2018}. When there are continuous
covariates, one can use the smoothed maximum score (SMS) estimator proposed
by \cite{horowitz_smoothed_1992} that employs a smooth approximation of the
original maximum score objective function. The SMS estimator is
asymptotically normally distributed and can have a faster rate of
convergence than the unsmoothed maximum score estimator. However, for
implementation of the SMS estimator, users have to choose tuning parameters
that might induce a smoothing bias which could be difficult to correct (\cite%
{kotlyarova_robust_2009}). We refer the reader to 
\citet[][Chapter
4]{horowitz_semiparametric_2009} for a detailed review on the theoretical
aspects of the maximum score and SMS estimators.

\section{An empirical model and parameters of interest}
We now present an empirical model in the context of happiness application. Suppose we are interested in using SWB data taken from two groups, say $A$ and $B$, to draw conclusions on whether people in group $A$ are happier than those in group $B$. Examples of group identities include gender, martial status, country, time etc. Previous analyses have been focusing on the mean as a statistic to compare happiness across groups. We will focus on the median and discuss analogous results for other quantiles. 

\subsection{Model}
Let observable data for an individual be $\left( Y,X,D\right) $ where $Y$%
\ denotes a reported happiness scale taking values from $\mathcal{Y}=\left\{
1,\ldots ,J\right\} $ for $J\geq 3$, $X$ is a vector of covariates other
than the group identity of interest, and the latter is denoted by a dummy variable $D$ that takes value $1$ for group $A$ and $0$
for group $B$. We assume $Y$ is derived from a threshold crossing model
based on latent happiness index, $H$, such that%
\begin{eqnarray}
Y &=&\left\{ 
\begin{array}{c}
1 \\ 
2 \\ 
\vdots  \\ 
J%
\end{array}%
\begin{array}{c}
\text{if }\gamma _{0}<H\leq \gamma _{1} \\ 
\text{if }\gamma _{1}<H\leq \gamma _{2} \\ 
\vdots  \\ 
\text{if }\gamma _{J-1}<H\leq \gamma _{J}%
\end{array}%
\right. \mathbf{,}  \label{pooled model} \\
\notag
\end{eqnarray}%
where
\begin{equation*}
H = X^{\top }\beta _{0}+\beta _{1}D+U
\end{equation*}
for some strictly increasing real thresholds $\left\{ \gamma _{j}\right\}
_{j=1}^{J-1}$ with $\gamma _{0}=-\infty ,\gamma _{J}=+\infty $, $\beta _{0}$
is a vector of parameters associated with $X$, $\beta _{1}$ is a scalar
parameter associated with $D$, $U$\ is an unobserved scalar accounting for
other factors, and the symbol $^{\top }$\ denotes the transpose.

The model described in this section is frequently used in
practice with an additional assumption on the distribution of $U$. For instance, a standardized probit model assumes $U|X,D \sim N(0,1)$ and a generalized probit model assumes $U|X,D \sim N(0,\sigma^{2}(X,D))$ where $\sigma^{2}(X,D)$ is specified up to some unknown parameters. In what follows, we shall focus on the use of the sign of $\beta _{1}$\ for ranking the
median across groups. Particularly, the definition of the median rank can be stated without specifying the distribution of $U|X,D$. However, as we shall see, some assumption on the distribution of $U|X,D$ is required to identify the median rank.

\subsection{Median Rank}

The parameter of interest for median comparison across groups with
observed characteristics $X$\ is: 
\begin{equation}
\lambda \left( X\right) :=Med\left( H|X,D=1\right) -Med\left( H|X,D=0\right)
.  \label{median comparison}
\end{equation}%
I.e. $\lambda \left( X\right) $\ is the difference between median levels of
happiness for individuals from groups $A$ and $B$ with the
same characteristics $X$. The median is equivariant to increasing transformation, so the signs for \ 
\begin{equation*}
\lambda \left( X;\tau \right) :=Med\left( \tau \left( H\right) |X,D=1\right)
-Med\left( \tau \left( H\right) |X,D=0\right) ,
\end{equation*}%
will be the same for all increasing function $\tau $. The sign of $\lambda
\left( X\right) $\ can thus be used to identify the median rank between $%
H|X,D=1$\ and $H|X,D=0$. We state this as a proposition.

\bigskip

\textsc{Proposition 1. }


The median rank of:

\begin{enumerate}
\item[(a)] $H|X,D=1$ is higher than $H|X,D=0$ if and only if $\lambda \left(
X\right) >0$;

\item[(b)] $H|X,D=1$ is lower than $H|X,D=0$ if and only if $\lambda \left(
X\right) <0$;

\item[(c)] $H|X,D=1$ is the same as $H|X,D=0$ if and only if $\lambda \left(
X\right) =0$.
\end{enumerate}

\bigskip 

Proposition 1 shows the sign of $\lambda \left( X\right) $\ can be used to
establish the median rank regardless whether $H$\ is homoskedastic or not,
and its validity does not require any knowledge of the distributional
assumption. 

We can impose some assumptions on the
distribution of $U|X,D$ for the identification of $\lambda \left( X\right) $. The simplest approach is to assume $U|X,D$ follows either a normal or logistic
distribution. Example 2 below illustrates this using a more general
probit model than Example 1.

\bigskip

\textsc{Example 2: }Consider model (\ref{pooled model}) where $U|X,D\sim
N\left( 0,\sigma ^{2}\left( X,D\right) \right) $ and $\gamma _{1}=0$ and $%
\gamma _{2}=1$. Then $H|X,D\sim N\left( X^{\top }\beta _{0}+\beta
_{1}D,\sigma ^{2}\left( X,D\right) \right) $ and $\lambda \left( X\right)
=\beta _{1}$. The sign of $\beta _{1}$\ determines the median happiness rank
between groups $A$ and $B$. This holds irrespective of the form of $\sigma ^{2}\left( X,D\right)$. 

\bigskip 

Estimation of $\lambda \left( X\right) $\ for probit and logit models can be
performed using standard statistical softwares. We emphasize, however, such parametric assumptions are not necessary
for identification and estimation. We will describe a semiparametric estimator in Section 4 where the only thing we assume about $U$ is a median restriction.

\subsection{$\protect\alpha $-Quantile Rank}

The previous discussion focuses on median rank. The median is a special case
of an $\alpha$-quantile with $\alpha =0.5$. More specifically, let us
denote the conditional $\alpha $-quantile of a continuous random variable $Z$\ given $W$ by 
$Q_{Z}\left( \alpha |W\right) $\ so that it satisfies,%
\begin{equation*}
\Pr \left( Z\leq Q_{Z}\left( \alpha |W\right) |W\right) =\alpha \text{
for any }\alpha \in \left( 0,1\right) .
\end{equation*}%
In happiness applications it is commonly assumed that $U$, and hence $H$, is a continuous random
variable. In this case $Q_{H}\left( \alpha |X,D\right) $\ is equal to the
inverse of the CDF of $H$ conditional on $\left( X,D\right) $\ evaluated at the quantile level $%
\alpha $. 

The parameter for performing $\alpha$-quantile comparison across groups with observed characteristics $X$, which is the counterpart to (\ref{median comparison}), is: 
\begin{equation*}
\lambda \left( \alpha |X\right) :=Q_{H}\left( \alpha |X,D=1\right)
-Q_{H}\left( \alpha |X,D=0\right) .
 \label{quantile comparison}
\end{equation*}%
I.e. $\lambda \left( \alpha |X\right) $\ is the difference between $\alpha$-%
quantile levels of happiness for individuals from groups $A$
and $B$ with the same characteristics $X$. Like the median, all quantiles are
equivariant to increasing transformations\footnote{%
That is, for any increasing function $\tau $, $Q_{\tau \left( H\right)
}\left( \alpha |X,D\right) =\tau \left( Q_{H}\left( \alpha |X,D\right)
\right) $.} property, so we know the signs of 
\begin{equation*}
\lambda \left( \alpha |X;\tau \right) :=Q_{\tau \left( H\right) }\left(
\alpha |X,D=1\right) -Q_{\tau \left( H\right) }\left( \alpha |X,D=1\right) ,
\end{equation*}%
will be invariant for every increasing function $\tau $. Thus,
the sign of $\lambda \left( \alpha |X;\tau \right) $\ can be used to
identify the $\alpha$-quantile rank between $H|X,D=1$\ and $H|X,D=0$. 

\bigskip 

\textsc{Proposition 2. } For any $\alpha\in(0,1)$, the $\alpha$-quantile rank of:

\begin{enumerate}
\item[(a)] $H|X,D=1$ is higher than $H|X,D=0$ if and only if $\lambda \left(
\alpha |X\right) >0$;

\item[(b)] $H|X,D=1$ is lower than $H|X,D=0$ if and only if $\lambda \left(
\alpha |X\right) <0$;

\item[(c)] $H|X,D=1$ is the same as $H|X,D=0$ if and only if $\lambda \left(
\alpha |X\right) =0$.
\end{enumerate}

\bigskip 

Similar to Proposition 1, quantile ranks is determined by the sign of $%
\lambda \left( \alpha |X\right) $\ without homoskedasticity or
distributional assumptions. 

In practice, we can use a linear quantile model for latent happiness $H
$\ to identify $\lambda \left( \alpha |X\right) $. For example, suppose we 
assume 
\begin{equation}
\label{eqn:quantile for H}
Q_{H}\left( \alpha |X,D \right)
=X^{\top }\beta _{0}\left( \alpha \right) +\beta _{1}\left( \alpha \right) D%
\text{ for any }\alpha \in \left( 0,1\right) \text{.}
\end{equation}

\noindent Then $\lambda \left( \alpha |X\right) =\beta _{1}\left( \alpha \right)$. Estimation of $\beta _{1}\left( \alpha \right)$ and $\beta _{0}\left( \alpha \right)$ is not straightforward even under parametric assumptions as these coefficients depend on the specification of heteroskedasticity of the distribution of $H$ conditional on $X$ and $D$. If we were able to observe $H$ and assume it has a continuous distribution, then we could estimate $\left( \beta _{0}\left( \alpha \right),\beta _{1}\left( \alpha \right) \right)$ using the quantile regression method of \cite{koenker_simple_1978}. However, such approach is not applicable in our case because the outcome $Y$ is discrete. In the next section, building on the ordinal median regressor estimator of \cite{lee_median_1992},  we will present an estimation approach that allows for estimating the quantile regression coefficients of (\ref{eqn:quantile for H}) in the context of discrete ordered response model (\ref{pooled model}).

\section{Estimating semiparametric ordinal median regression model through mixed integer optimization%
}

We begin this section by outlining the main challenges relating to the
estimation of semiparametric median and quantile regressions for ordinal
outcomes. We propose that a mixed integer optimization procedure
can be useful in this setting and propose an estimation procedure for them. The details on mixed integer optimization can be found in the appendix. As in the previous section, we will focus on the median case and discuss the cases for other quantiles afterwards.

\subsection{A least absolute deviation estimation problem}

We begin with the following model of semiparametric median regression for
ordinal outcomes: 
\begin{eqnarray}
Y &=&\sum\nolimits_{j=1}^{J}j\times \mathbf{1}\left\{ \gamma _{j-1}<H\leq
\gamma _{j}\right\} ,  \label{model 1a} \\
H &=&X^{\top }\theta +U,  \label{median model 1a}
\end{eqnarray}%
where the covariate vector $X\in\mathbb{R}^{p+1}$ subsumes group dummy variables. Equation (\ref{model 1a}) amounts to rewriting (\ref{pooled model}) using
indicator functions. Let $\gamma _{0}=-\infty ,\gamma _{J}=+\infty $\ and $%
\gamma :=\{\gamma _{j}\}_{j=1}^{J-1}$ denotes the vector of other threshold
parameters that are strictly increasing. The parameters of the model are $\left( \theta ,\gamma \right) $.

For presenting our estimation procedure, it will be convenient to write $%
X=(X_{1},\widetilde{X})$, where $X_{1}\in \mathbb{R}$ is the first element
of $X$\ and $\widetilde{X}\in \mathbb{R}^{p}$ is a vector containing the remaining
components of $X$. Correspondingly, we let $\theta =(\theta _{1},\beta )$ so that $%
X^{\top }\theta =\theta _{1}X_{1}+\widetilde{X}^{\top }\beta $. Since parameters of the
ordinal choice model (\ref{model 1a})-(\ref{median model 1a}) are not identified without location and
scale normalizations, we assume $X$\ does not contain a constant (location
normalization) and $\left\vert \theta _{1}\right\vert =1$ (scale
normalization). There are other ways to normalize. For example, another convenient
choice is to set $\gamma _{1}=0$ and $\gamma _{2}=1$ as done in Example 1. We refer the reader to \cite{greene_hensher_2010} for further discussions on normalization in ordered choice models.

To estimate the median semiparametrically we assume that $Med(U|X)=0$ for all $X$%
.\ This allows for nonparametric specification of the distribution of
latent unobservable $U$, which admits general unknown form of
heteroskedasticity conditional on the covariates. In particular, it can be shown the zero median condition implies that
\begin{equation}
\label{Sign Match}
P(Y \leq j|X) \geq 0.5 \iff \gamma_j \geq X^{\top }\theta  ,
\end{equation}%
which subsequently yields,

\begin{equation}
\label{Med(Y|X)}
Med(Y|X)=\sum\nolimits_{j=1}^{J}j\times \mathbf{1}\left\{ \gamma
_{j-1}<X^{\top }\theta \leq \gamma _{j}\right\} .
\end{equation}%
We want to highlight that the previous two equations do not depend on the variance or other distributional knowledge about $U|X$. This feature is attractive because we can then use (\ref{Med(Y|X)}) to estimate the median without any risk of misspecifying aspects about the distribution of $U|X$. It is in this sense that the semiparametric estimator is robust. 

Let $\Theta \subset \mathbb{R}^{p+J-1}$\ denote the parameter space containing $\left( \beta ,\gamma \right) $. We use $\left( b,c\right) $ to denote a generic point of $\Theta $. Then, given a random sample $\left( Y_{i},X_{i}\right) _{i=1}^{n}$, letting $%
c_{0}=-\infty $ and $c_{J}=\infty $, consistent estimator $(\widehat{\theta }%
_{1},\widehat{\beta },\widehat{\gamma })$ of $(\theta _{1},\beta ,\gamma )$
can be obtained as a solution to the following least absolute deviation
(LAD) estimation problem\footnote{%
We refer the reader to \cite{lee_median_1992} for a comprehensive discussion of this LAD estimator.}:%
\begin{equation}
\min_{(a,b,c)\in \{1,-1\}\times \Theta }\sum\nolimits_{i=1}^{n}\left\vert
Y_{i}-\sum\nolimits_{j=1}^{J}j\times \mathbf{1}\left\{ c_{j-1}<aX_{1i}+%
\widetilde{X}_{i}^{\top }b\leq c_{j}\right\} \right\vert .
\label{LAD formulation 1}
\end{equation}%
The LAD estimator that solves the minimization problem in (\ref{LAD formulation 1}) is a generalization of
the maximum score estimation approach of \cite{manski_semiparametric_1985}
for the binary response model to that for the discrete ordered response
case. As in the problem of maximum score estimation, the objective function
in (\ref{LAD formulation 1}) is piecewise constant with numerous local minima. Thus, although this LAD estimator is theoretically well-defined, finding a global solution in practice is known to be a notoriously difficult task.

We propose an algorithm which enables exact computation of
a global solution to the LAD problem (\ref{LAD formulation 1}) using mixed integer linear programming. The idea is to turn the difficult optimization problem with a non-convex objective function over a convex domain ($\Theta $) to an optimization problem with a linear objective function over a domain that embodies integrality restrictions, which means the parameters are passed on as a set of linear inequality constraints that need to be satisfied. Specifically, we show that the integer values to be
considered are finite, which can then be efficiently searched using modern mixed integer optimization (MIO) methods. This is the insight that \cite{florios_exact_2008} have exploited to compute Manski's maximum score estimator. The details of the mixed integer optimization procedures that can be used to estimate the semiparametric median as well as other quantiles are given in the Appendix.

\subsection{Parametric and semiparametric estimation}

The theoretical advantages of the semiparametric estimator is that the
results will be robust to model misspecification, particularly on the form
of heteroskedasticity and distribution of $U$ conditional on $X$. In particular, if the
functional form of the variance of $U$ given $X$ is misspecified in probit or logit
models, or if the conditional distribution of $U$ given $X$ is in fact not normal or logistic
respectively, these parametric estimators are inconsistent.\ On the other hand, if
the parametric assumptions are correct then the corresponding parametric estimator will be asymptotically more efficient than the semiparametric estimator in terms of having smaller
asymptotic variance.

There are other considerations to bear in mind when using the semiparametric estimator. First, the researcher has to choose the variable $X_1$ that is associated with $\theta_1$, which we normalize to be -1 or 1, carefully. The chosen $X_1$ is assumed a priori to have non-zero effect on latent happiness though the sign of its effect, positive or negative, can be estimated. Ideally  $X_1$ should also have a rich support as it facilitates parameter identification.\footnote{Standard sufficient conditions for identification in maximum score type estimation problems assume one of the covariates has full support on $\mathbb{R}$ conditional on other explanatory variables. The full support condition is, however, not necessary. It can be reduced to bounded or even finite support as explained in \cite{manski_identification_1988} and \citet[][Chapter 4]{horowitz_semiparametric_2009}.} 
On the practical front, estimating the semiparametric median
regression is also computationally more involved than estimating ordered probit or logit
models. In particular, despite using an efficient solver, it generally can take a long time to solve an MIO problem and the computational time increases with the sample size. This latter point can be seen by inspecting equation (\ref{LAD MIO formulation}) in the appendix where it is apparent that the number of the optimizing integer variables grows with the sample size. In contrast, it is much easier to compute the maximum likelihood estimators for the logit and probit models. Inference on parametric models is also relatively straightforward. For instance, the Stata commands mentioned earlier automatically provide standard errors and confidence intervals for the median estimators. Inference based on the semiparametric estimator here can be performed by resampling, which further exacerbates the computational cost.

By design, it is more difficult to identify and estimate the model without additional parametric structures. Yet the semiparametric approach can deliver more robust results against model misspecifications. 
Our view is that parametric and semiparametric estimators should be complementary rather than used as a substitute for each other in conducting empirical studies. 

\subsection{Extension to quantile regressions for ordinal outcomes}
We now extend the ordinal median regression approach of Section 4.1 to the setting of estimation at other quantiles. Let us consider a linear quantile model for the latent variable $H$ in (\ref{model 1a}). Specifically, using the notation introduced in Section 3.3, we assume that for $\alpha \in (0,1) $, $Q_{H}\left( \alpha |X\right)=X^{\top }\theta $. This assumption implies that there is a residual $U$ such that (\ref{median model 1a}) holds with $Q_{U}\left( \alpha |X\right)=0$. In this case it can be shown that, 
\begin{equation*}
\label{Q(Y|X)}
Q_{Y}\left( \alpha |X\right)=\sum\nolimits_{j=1}^{J}j\times \mathbf{1}\left\{ \gamma
_{j-1}<X^{\top }\theta \leq \gamma _{j}\right\},
\end{equation*}    
which encompasses (\ref{Med(Y|X)}) as a special case where $\alpha=0.5$. \par
Using the same location and scale normalizations as in Section 4.1, we aim to estimate $\theta=(\theta_{1},\beta)$ where we can write $X^{\top }\theta=\theta_{1}X_{1}+\tilde{X}^{\top}\beta$ with $\vert\theta_{1}\vert=1$. In particular, analogously to (\ref{LAD formulation 1}), we can estimate $(\theta_{1},\beta,\gamma)$ by solving the following minimization problem:
\begin{equation}
\min_{(a,b,c)\in \{1,-1\}\times \Theta }\sum\nolimits_{i=1}^{n}\rho_{\alpha}\left(
Y_{i}-\sum\nolimits_{j=1}^{J}j\times \mathbf{1}\left\{ c_{j-1}<aX_{1i}+%
\widetilde{X}_{i}^{\top }b\leq c_{j}\right\} \right),
\label{QR formulation 1}
\end{equation}%
where $\rho_{\alpha}$ is the \textit{check function} used in quantile regression (\cite{koenker_simple_1978}) defined as $\rho_{\alpha}(u):=u\left(\alpha-1\left\{u\leq0\right\}\right)$ for any $u\in\mathbb{R}$. Indeed, the median estimator that minimize (\ref{LAD formulation 1}) coincides with that of (\ref{QR formulation 1}) when $\alpha = 0.5$.\par
It is not a simple task to solve (\ref{QR formulation 1}) for the same reason that makes solving (\ref{LAD formulation 1}) difficult. Conveniently, the computational strategy we proposed based on  mixed integer linear programming (MILP) to estimate the median can readily be adapted to estimate other quantiles. 
Thus, the semiparametric ordinal quantile regression estimator can be effectively computed through the method of mixed integer optimization for any quantile level.

\section{Revisiting Happiness Equation}

To demonstrate the practical usefulness of our approaches, we apply them to estimate a standard happiness equation for studying correlation between individual well-being and demographic and socioeconomic characteristics (see e.g. \cite{blanchflower_well-being_2004}). Previous studies have focused on understanding how certain socio-demographic characteristics, e.g. unemployment and martial status, are related to the average well-being. We are interested in learning how these factors affect the median and other quantiles of happiness distribution. More importantly, we are interested in whether the structure of the happiness equation remains qualitatively similar when the median instead of the mean is the outcome of interest. In particular, we consider ordered probit and logit models, where the underlying happiness distribution is assumed to be homoskedastic as well as heteroskedastic. We also estimate the semiparametric quartiles of the happiness distribution, viz. the 25th, 50th, and 75th percentiles, to investigate heterogeneous effects at different parts of conditional happiness distribution. 

We use bi-annual data taken from the US General Social Survey (GSS) between the years 2000 and 2018 inclusive. It consists of 19,275 observations. The GSS happiness variable comes from respondents being asked whether they are ``1. not too happy'', ``2. pretty happy'', or ``3. very happy''. The explanatory variables for the happiness equations are income, age, age-squared, sex, level of education, marital status, race, employment status, and time fixed effects. We use the logarithm of equivalence scale adjusted household income. A set of education dummies includes indicators for those who left high school, completed a bachelor and had a graduate degree, with high school/junior college being the omitted category. The marital status dummies include married, divorced, widowed, with the omitted category being never married. Race dummies include black and other race, white is used as the omitted category. Finally, employment status dummies include unemployed, non-labour status, and the omitted category is for those that work fulltime or parttime.

We estimate our parametric models under normal and logistic distributions using Stata. For the standardized probit and logit models, which assume homoskedasticity, we use the commands \texttt{oprobit} and \texttt{ologit} respectively. For the generalized ordered probit and logit models that allow for parametric forms of heteroskedasticity we use the \texttt{oglm} command.\footnote{Due to the many explanatory variables involved we specify the heteroskedastic function parametrically as the nonparametric approach in \cite{oparina_analyzing_2021} does not work well in this scenario. Specifically, the nonparametric approach requires nonparametric estimation of the conditional probabilities for different happiness levels. We would have very few observations in some cells leading to imprecise estimates because we have many discrete explanatory variables to condition on.} Specifically, the vector of covariates in the latter models does not contain a constant, as a location normalization, and the conditional variance function is an exponential function that contains the same vector of covariates to incorporate scale normalization.\footnote{Using the notation in (\ref{median model 1a}), we use \texttt{oglm} to estimate the generalized ordered probit under the assumption that $H=X^{\top
}\theta +U$ where $U=\sigma ( X^{\top }\delta ) \varepsilon $ with $\sigma^{2}(t)=e^t$ and $%
\varepsilon \sim N\left( 0,1\right) $ that is independent of $X$. We
estimate the generalized ordered logit analogously where $\varepsilon $ would then follow a standard logistic
distribution. Further details on \texttt{oglm} can be found in \cite{williams_fitting_2010}.}

To estimate the semiparametric median and quartiles, we impose the normalizations  described in Section 4.1. We do not have the constant term in the regression and the magnitude of the coefficient on income is set to one. We note that our income variable has a rich support\footnote{The GSS collected household income data in 23 to 26 bands (depending on the year). We computed their mid-points and debased them accordingly. Subsequently, the income variable in our sample can take 246 possible values. Since we used equivalence scale adjusted household income, it takes 1425 possible values altogether.}. Our semiparametric estimates are obtained using the MILP procedure described in the appendix. Specifically, we used the MATLAB implementation of the Gurobi Optimizer to solve the MILP problems.\footnote{The codes and data used for obtaining the results in our paper are publicly available from the following repository \url{https://data.mendeley.com/datasets/dj5mpwypsc}.} 

Table 1 presents the ordered logit and probit estimation results. The top half of the table shows those estimates that are associated with the median, which is the same as the mean in these instances, and the bottom half are concerned with the variance estimates. We only report estimates for the socioeconomic factors. Time fixed effects, which are typically not the focus when analyzing the happiness equation, are omitted\footnote{For a particular model, the time effects are statistically significant at $1\%$ for some years but not significant at $10\%$ for some other years; there is no clear pattern for statistical significance across years. However, the signs of the effects and significance results for each year are the same in all models.} for the sake of space. 

The median/mean results for the socioeconomic effects are in line with previous findings often reported in the well-being literature. For example, respondents with higher income report higher levels of well-being (e.g., \citet{easterlin_does_1974}, \citet{ferrer-i-carbonell_income_2005}); well-being follows a U-shape throughout one’s life (e.g., \citet{blanchflower_is_2008}) where the convexity in age has been attributed to the midlife nadir (e.g. see \cite{cheng_longitudinal_2017}); women are happier than men (\citet{stevenson_paradox_2009}); university graduates are happier than those who do not have a degree (\citet{blanchflower_well-being_2004}); being married is positively related with well-being, while being black or of other race, unemployed or not in labour force has negative correlation (e.g., \citet{easterlin_explaining_2003}). We find almost no qualitative difference in the median estimates from the homoskedastic model and the heteroskedastic one despite there being evidence for heteroskedasticity. On the latter, we find income, low educational attainments, race, and employment status have significant effects on the variance of conditional happiness distribution of respective groups. Formal tests, based on the Wald statistic, also reject the null hypothesis of homoskedasticity\footnote{The null hypothesis corresponds to the setting where all the variance coefficients in the conditional variance function are zero.} with p-values below $0.01$. The probit and logit results described above hold almost identically except that the effect of the \textit{other race} status is significant at $10\%$ for the heteroskedastic logit model but not significant at that level in all other cases. The fact that we find evidence of income affecting happiness, even in parametric models where the results may be biased due to misspecification, is a useful precursor for semiparametric estimation. This is because in the semiparametric model we assume the income coefficient is non-zero when we impose the scale normalization through this parameter.

Table 2 tabulates the median estimates of parameters and their confidence intervals from the generalized probit and logit models and the semiparametric counterpart. We note that the parametric estimates in Table 2 differ from those in Table 1 because we divide the parametric estimates by their respective estimated coefficient of income to facilitate comparisons with the semiparametric estimates. The reported semiparametric estimation results correspond to the set of estimates obtained when the sign of the income coefficient is positive, which is the sign that minimizes the LAD problem, so it is consistent with the positive income effects seen in the parametric model. The $95\%$ confidence intervals for the parametric and semiparametric estimators are computed using normal approximation and $m$ out of $n$ bootstrap respectively. 
We note that for Tables 2 and 3, we only signify statistical significance at $5\%$ level as the only confidence intervals constructed for our semiparametric estimators are the $95\%$ confidence intervals.

The most striking aspect from Table 2 is that all parametric and semiparametric estimates have the same sign for all socioeconomic variables. The insignificant/significant effects at the 5\% level also coincide in most of these cases. The semiparametric model suggests more pronounced effects, in terms of statistical significance, for being \textit{widowed} or \textit{other race} than the parametric models, and finds the \textit{unemployed} status to be negative but insignificant. These differences may be due to the symmetric distributional assumption imposed by the parametric models.

To better understand how well-being is related with individual characteristics across happiness distributions, Table 3 provides the estimates across the first (25th percentile), second (50th percentile) and third (75th percentile) quartiles. Most socioeconomic variables affect happiness across the three quartiles in similar ways if we look at the signs of the estimates. There are, however, some differences across quartiles in terms of statistical significance for some factors, but none in which the signs differ for different quartiles that are both statistically significant. Two interesting differences, using the median as the benchmark, are: the negative effect from being \textit{unemployed}, which is not significant for the median, is significant for the lower and upper quartiles; and, the significant negative impact for being \textit{black} or \textit{other race} also stand in contrast with the non-significant effects for the lower quartile (negative) and upper quartile (positive). Other than these we find the factors that drive happiness at the 50th and 75th percentiles to be the same. In contrast, a particular pattern we observe is that, at the 25th percentile of happiness distribution, several drivers (\textit{female}, \textit{bachelor}, \textit{widowed}, \textit{black}, \textit{other race}) of happiness that are significant at higher percentiles become insignificant. 

Taken together, Tables 1 - 3 show our results largely align with previous findings in the well-being literature, which will be reassuring for many researchers working in this field. This is particularly clear for median regression analysis, and also true to a large extent at other quartiles of the happiness distribution. We summarize our results as follows.
\begin{itemize}
  \item The median estimates using ordered probit and logit models re-affirm the conventional effects that socioeconomic factors in happiness equation have on median happiness. 
  \item Median estimates from the semiparametric model are qualitatively very similar to those from the parametric models. This suggests the symmetry assumption that underlie probit and logit models does not drive the results on how most of the socioeconomic factors affect the median of happiness distribution. 
  \item Many of the socioeconomic factors that drive the median happiness have the same effect for the lower and upper quartiles. However, there are some interesting differences; for example, fewer factors tend to matter to the 25th percentile of the happiness distribution relative to the 50th and 75 percentiles.
\end{itemize}

\newpage

\begin{table}[H]
\renewcommand{\arraystretch}{0.8}
\begin{centering}
\addtolength{\tabcolsep}{7pt} 
\resizebox{\textwidth}{!}{
\begin{tabular}{lS[table-format=-1.3]cS[table-format=-1.3]cS[table-format=-1.3]cS[table-format=-1.3]c}
\hline
\hline
                     & \multicolumn{4}{c}{Homoskedastic estimates}               & \multicolumn{4}{c}{Heteroskedastic estimates}          \\
                     & \multicolumn{2}{c}{Logit}    & \multicolumn{2}{c}{Probit} & \multicolumn{2}{c}{Logit} & \multicolumn{2}{c}{Probit} \\
                     & \multicolumn{1}{c}{coef.}           & \multicolumn{1}{c}{SE}          & \multicolumn{1}{c}{coef.}          & \multicolumn{1}{c}{SE}           & \multicolumn{1}{c}{coef.}         & \multicolumn{1}{c}{SE}           & \multicolumn{1}{c}{coef.}          & \multicolumn{1}{c}{SE}           \\\hline
income               & 0.245***        & (0.021)    & 0.138***      & (0.012)    & 0.334***     & (0.040)    & 0.179***      & (0.020)    \\
age                  & -0.061***       & (0.006)    & -0.034***     & (0.004)    & -0.078***    & (0.010)    & -0.043***     & (0.005)    \\
age squared          & 0.001***        & (0.000)    & 0.000***      & (0.000)    & 0.001***     & (0.000)    & 0.000***      & (0.000)    \\
female               & 0.103***        & (0.033)    & 0.061***      & (0.019)    & 0.137***     & (0.044)    & 0.076***      & (0.025)    \\
left high school     & -0.079          & (0.059)    & -0.048        & (0.033)    & -0.096       & (0.078)    & -0.052        & (0.042)    \\
bachelor             & 0.254***        & (0.044)    & 0.154***      & (0.026)    & 0.313***     & (0.062)    & 0.184***      & (0.035)    \\
graduate             & 0.216***        & (0.055)    & 0.134***      & (0.032)    & 0.261***     & (0.073)    & 0.155***      & (0.041)    \\
married              & 0.906***        & (0.047)    & 0.521***      & (0.027)    & 1.119***     & (0.111)    & 0.625***      & (0.058)    \\
divorced             & -0.037          & (0.056)    & -0.022        & (0.031)    & -0.047       & (0.072)    & -0.021        & (0.039)    \\
widowed              & -0.106          & (0.084)    & -0.056        & (0.047)    & -0.151       & (0.109)    & -0.077        & (0.059)    \\
black                & -0.125**        & (0.051)    & -0.078***     & (0.029)    & -0.179***    & (0.069)    & -0.098***     & (0.037)    \\
other race           & -0.079          & (0.063)    & -0.050        & (0.036)    & -0.141*      & (0.083)    & -0.071        & (0.046)    \\
unemployed           & -0.502***       & (0.073)    & -0.289***     & (0.041)    & -0.634***    & (0.110)    & -0.349***     & (0.058)    \\
non-lf status        & -0.096**        & (0.044)    & -0.061**      & (0.025)    & -0.115**     & (0.056)    & -0.063**      & (0.031)    \\
\hline
\multicolumn{2}{l}{Variance estimates} &            &               &            &              &            &               &            \\
\hline
income               &                 &            &               &            & -0.059***    & (0.013)    & -0.049***     & (0.011)    \\
age                  &                 &            &               &            & 0.004        & (0.004)    & 0.002         & (0.003)    \\
\multicolumn{2}{l}{age squared}        &            &               &            & 0.000        & (0.000)    & 0.000         & (0.000)    \\
female               &                 &            &               &            & -0.032       & (0.021)    & -0.024        & (0.019)    \\
\multicolumn{2}{l}{left high school}   &            &               &            & 0.120***     & (0.033)    & 0.109***      & (0.031)    \\
bachelor             &                 &            &               &            & -0.015       & (0.028)    & 0.000         & (0.025)    \\
graduate             &                 &            &               &            & 0.001        & (0.037)    & 0.018         & (0.033)    \\
married              &                 &            &               &            & -0.009       & (0.028)    & 0.035         & (0.025)    \\
divorced             &                 &            &               &            & 0.023        & (0.032)    & 0.024         & (0.029)    \\
widowed              &                 &            &               &            & -0.002       & (0.047)    & 0.002         & (0.043)    \\
black                &                 &            &               &            & 0.140***     & (0.029)    & 0.126***      & (0.026)    \\
other race           &                 &            &               &            & 0.084**      & (0.037)    & 0.077**       & (0.034)    \\
\multicolumn{2}{l}{unemployed}         &            &               &            & 0.103**      & (0.041)    & 0.086**       & (0.038)    \\
\multicolumn{2}{l}{non-lf status}      &            &               &            & 0.100***     & (0.026)    & 0.089***      & (0.024)    \\
\hline
cut 1                & -3.145***       & (0.147)    & -1.826***     & (0.084)    & -4.056***    & (0.383)    & -2.232***     & (0.189)    \\
cut 2                & -0.151***       & (0.144)    & -0.058***     & (0.083)    & -0.197       & (0.186)    & -0.071        & (0.103)    \\
\hline
Observations         &  \multicolumn{1}{c}{19,275} & & \multicolumn{1}{c}{19,275} &  &\multicolumn{1}{c}{19,275} & & \multicolumn{1}{c}{19,275} &            \\ 
\hline
\hline
\end{tabular}
}
\caption{Parametric median estimated estimates}
\end{centering}
\medskip {\small 
* p$<$0.10, {**} p$<$0.05, *** p$<$0.01 \\
\textit{income} is the standardized logarithm of equivalence scale adjusted household income; \textit{%
age} is the respondent's age; \textit{age squared} is age squared; \textit{female} is a dummy for being female; \textit{left high school}, \textit{bachelor}, \textit{graduate} are dummy variables indicating respective educational attainments; \textit{married}, \textit{divorced}, \textit{widowed} indicate marital status; \textit{black} and \textit{other race} are dummy variables for race with \textit{white} being a missing category; \textit{unemployed} and \textit{non-lf status} are dummies indicating respectively whether the respondent is currently unemployed and has a non-labour force status, including retired, student, keeping home and others.}
\end{table}

\begin{table}[H]
\renewcommand{\arraystretch}{0.8}
\begin{centering}
\addtolength{\tabcolsep}{-6pt} 
\resizebox{\textwidth}{!}{
\begin{tabular}{l|*{3}{S[table-format=-2.3,table-space-text-post=***]}|*{3}{S[table-format=-2.3,table-space-text-post=***]}|*{3}{S[table-format=-2.3,table-space-text-post=***]}}
 \hline
 \hline
 & \multicolumn{6}{c}{Parametric est.} & \multicolumn{3}{c}{Semiparametric est.} \\
 & Logit & \multicolumn{2}{c}{95\% Conf. Interval} & Probit & \multicolumn{2}{c}{95\% Conf. Interval} &  & \multicolumn{2}{c}{95\% Conf. Interval} \\ \hline
age              & -0.232**  & -0.289  & -0.176  & -0.237**  & -0.296  & -0.178  & -0.238**  & -0.309  & -0.185  \\
age squared      & 0.002**   & 0.002   & 0.003   & 0.002**   & 0.002   & 0.003   & 0.002**   & 0.002   & 0.003   \\
female           & 0.412**   & 0.161   & 0.662   & 0.425**   & 0.164   & 0.687   & 0.421**   & 0.091   & 0.752   \\
left high school & -0.287    & -0.751  & 0.177   & -0.288    & -0.759  & 0.182   & -0.221    & -0.817  & 0.375   \\
bachelor         & 0.937**   & 0.552   & 1.322   & 1.029**   & 0.617   & 1.441   & 1.008**   & 0.668   & 1.541   \\
graduate         & 0.781**   & 0.328   & 1.233   & 0.866**   & 0.379   & 1.352   & 0.858**   & 0.254   & 1.479   \\
married          & 3.349**   & 2.665   & 4.032   & 3.486**   & 2.772   & 4.200   & 2.922**   & 2.591   & 3.683   \\
divorced         & -0.141    & -0.565  & 0.283   & -0.117    & -0.539  & 0.304   & -0.533    & -1.076  & 0.010   \\
widowed          & -0.453    & -1.093  & 0.187   & -0.432    & -1.072  & 0.208   & -1.685**  & -2.502  & -0.868  \\
black            & -0.536**  & -0.942  & -0.130  & -0.544**  & -0.956  & -0.132  & -0.618**  & -1.140  & -0.096  \\
other race       & -0.423    & -0.911  & 0.065   & -0.397    & -0.901  & 0.106   & -1.410**  & -2.048  & -0.822  \\
unemployed       & -1.897**  & -2.548  & -1.246  & -1.949**  & -2.605  & -1.293  & -0.570    & -1.414  & 0.274   \\
non-lf status    & -0.343**  & -0.685  & -0.002  & -0.354    & -0.710  & 0.002   & -0.311    & -0.762  & 0.135   \\
\hline
cut 1            & -12.140** & -14.245 & -10.035 & -12.448** & -14.643 & -10.253 & -10.996** & -13.069 & -10.009 \\
cut 2            & -0.591    & -1.677  & 0.495   & -0.395    & -1.529  & 0.739   & -0.967    & -2.277  & 0.500   \\ \hline
Observations & \multicolumn{1}{c}{19,275} & & & \multicolumn{1}{c}{19,275} & & & \multicolumn{1}{c}{19,275} & & 
 \\\hline\hline 
\end{tabular}
}
\caption{Parametric and semiparametric estimates}
\end{centering}
\medskip {\small 
{**} p$<$0.05 \\
\textit{%
age} is the respondent's age; \textit{age squared} is age squared; \textit{female} is a dummy for being female; \textit{left high school}, \textit{bachelor}, \textit{graduate} are dummy variables indicating respective educational attainments; \textit{married}, \textit{divorced}, \textit{widowed} indicate marital status; \textit{black} and \textit{other race} are dummy variables for race with \textit{white} being a missing category; \textit{unemployed} and \textit{non-lf status} are dummies indicating respectively whether the respondent is currently unemployed and has a non-labour force status, including retired, student, keeping home and others.}
\end{table}

\begin{table}[H]
\renewcommand{\arraystretch}{0.8}
\begin{centering}
\addtolength{\tabcolsep}{-6pt} 
\resizebox{\textwidth}{!}{\begin{tabular}{l|*{3}{S[table-format=-2.3,table-space-text-post=***]}|*{3}{S[table-format=-2.3,table-space-text-post=***]}|*{3}{S[table-format=-2.3,table-space-text-post=***]}}
 \hline
 \hline
 & \multicolumn{3}{c|}{25 quantile} & \multicolumn{3}{c|}{Median} & \multicolumn{3}{c}{75 quantile} \\
 & \multicolumn{1}{c}{coef} & \multicolumn{2}{c|}{95\% Conf. Interval} & \multicolumn{1}{c}{coef} & \multicolumn{2}{c|}{95\% Conf. Interval} & \multicolumn{1}{c}{coef} & \multicolumn{2}{c}{95\% Conf. Interval} \\ \hline
age              & -0.260** & -0.333  & -0.212 & -0.238**  & -0.309  & -0.185  & -0.198** & -0.264  & -0.135 \\
age squared      & 0.002**  & 0.002   & 0.003  & 0.002**   & 0.002   & 0.003   & 0.002**  & 0.002   & 0.003  \\
female           & 0.283    & -0.046  & 0.604  & 0.421**   & 0.091   & 0.752   & 0.694**  & 0.363   & 1.025  \\
left high school & -0.675** & -1.271  & -0.079 & -0.221    & -0.817  & 0.375   & 0.501    & -0.095  & 1.097  \\
bachelor         & 0.296    & -0.145  & 0.782  & 1.008**   & 0.668   & 1.541   & 0.731**  & 0.230   & 1.259  \\
graduate         & 1.524**  & 1.171   & 2.145  & 0.858**   & 0.254   & 1.479   & 0.860**  & 0.239   & 1.481  \\
married          & 2.087**  & 1.546   & 2.646  & 2.922**   & 2.591   & 3.683   & 2.737**  & 2.358   & 3.496  \\
divorced         & 0.027    & -0.516  & 0.570  & -0.533    & -1.076  & 0.010   & 0.405    & -0.138  & 0.948  \\
widowed          & -0.701   & -1.518  & 0.115  & -1.685**  & -2.502  & -0.868  & -1.548** & -2.364  & -0.731 \\
black            & -0.503   & -1.025  & 0.019  & -0.618**  & -1.140  & -0.096  & 0.251    & -0.271  & 0.773  \\
other race       & -0.165   & -0.803  & 0.473  & -1.410**  & -2.048  & -0.822  & 0.332    & -0.306  & 0.970  \\
unemployed       & -2.350** & -3.194  & -1.506 & -0.570    & -1.414  & 0.274   & -0.932** & -1.776  & -0.088 \\
non-lf status    & -0.298   & -0.747  & 0.142  & -0.311    & -0.762  & 0.135   & -0.415   & -0.866  & 0.015  \\ \hline
cut 1            & -8.591** & -10.363 & -7.564 & -10.996** & -13.069 & -10.009 & -9.483** & -11.263 & -7.319 \\
cut 2            & 0.176    & -1.048  & 1.643  & -0.967    & -2.277  & 0.500   & -2.775** & -4.224  & -1.308 \\
\hline
Observations & \multicolumn{1}{c}{19,275} & & & \multicolumn{1}{c}{19,275} & & & \multicolumn{1}{c}{19,275} & & 
 \\\hline\hline 
\end{tabular}}
\caption{Semiparametric quantile estimates}
\end{centering}
\medskip {\small 
{**} p$<$0.05 \\
 \textit{age} is the respondent's age; \textit{age squared} is age squared; \textit{female} is a dummy for being female; \textit{left high school}, \textit{bachelor}, \textit{graduate} are dummy variables indicating respective educational attainments; \textit{married}, \textit{divorced}, \textit{widowed} indicate marital status; \textit{black} and \textit{other race} are dummy variables for race with \textit{white} being a missing category; \textit{unemployed} and \textit{non-lf status} are dummies indicating respectively whether the respondent is currently unemployed and has a non-labour force status, including retired, student, keeping home and others.}
\end{table}

\section{Conclusion}

A group ranking of ordinal outcomes is identified only when the ranking order is invariant across all increasing transformations on the ordinal variables. For the mean ranking, this invariance is equivalent to there being a first order stochastic dominance (FOSD) relation between the variables across groups. The usefulness of the probit and logit based mean ranking for discrete ordinal outcomes has in particular been put under question, as illustrated by \cite{bond_sad_2019}, because FOSD in this setting requires the model to be homoskedastic, otherwise the mean rank is not identified.

In this paper we propose focusing on the median as a pragmatic alternative to the mean. Firstly, the median rank of ordinal outcomes can be identified even when the mean rank is not. Secondly, probit and logit based median ranks can be identified by the conditional means of latent variables of these parametric models and can hence be easily estimated using standard statistical softwares. Thirdly, we also propose a mixed integer optimization procedure to perform median regression in a semiparametric ordered response model that can accommodate unknown distribution of the latent unobservable. Furthermore, our mixed integer optimization procedure can be used to estimate other quantiles of the distribution in addition to the median. Quartile ranks are also identified under weak conditions similar to the median. Estimates of quantile regressions can be useful for researchers to understand effects of happiness driving factors across different parts of the happiness distribution.

In our empirical study, we revisit the happiness equation for the US using the GSS data. We find that median estimates of the ordered probit and logit results are qualitatively very similar to the semiparametric ones. These results, which are in line with the conventions in the happiness literature, suggest that structures, e.g. symmetry, imposed by familiar parametric models have limited role in determining how different socioeconomic factors affect median of the happiness distribution. While many qualitative results of the estimated semiparametric median are also found in the estimation of the lower and upper quartiles of the happiness distribution, there are some notable differences. Compared to the higher quartiles, the happiness distribution at the lower quartile seem to be affected by fewer factors. For example, being widowed, black or other race seem to matter less. But having dropped out of high school or being unemployed seem to have particularly high negative association with happiness in the lower quartile. Thus, a policy maker who wants to improve welfare of people at lower quantiles may want to focus on these factors for further investigations. 

The model we consider in the paper is applicable in the context with single or repeated cross-sectional data. There are also many well-being applications that use panel data. The median in these models can be well estimated with a suitable procedure. For example, consider a panel probit or logit model of happiness with fixed
effects that satisfies: $E \left[ H_{it}|X_{it},\mu _{i} \right] =  X_{it}^{\top }\beta+\mu _{i},$ where the indices $i$ and $t$ denote individual and time
respectively, and $\mu _{i}$\ denotes the unobserved individual fixed
effect. By symmetry, the conditional median of $H_{it}$ given $\left( X_{it} , \mu_i \right)$ is also $X_{it}^{\top }\beta + \mu _{i}$. However, it is well-known that $\left( \beta,\mu _{i} \right)$ cannot be consistently estimated by maximum likelihood (ML) with fixed time periods $T$ due to the incidental parameter problem. Applications often focus on $ \beta$ and the finite sample bias of ML estimators can be substantial when $T$ is small. A popular way to deal with this in the econometrics literature is to conduct a bias-correction procedure. E.g., \cite{hahn_jackknife_2004} and \cite{bester_penalty_2012} provide methods to do this for general non-linear parametric panel data models that include ordered probit and logit. Another approach, which is based on dichotomization (\citet{chamberlain_analysis_1980}), specific for an ordered logit model can yield a consistent estimator of $\beta$ with finite $T$. Dichotomization-based estimators have been quite popular in well-being applications. We refer the reader to \cite{baetschmann_consistent_2015} for an examples of such estimators that include a consistent and efficient version. All these estimation methods are viable options for estimating the conditional median of probit and logit models with fixed effects. Semiparametric estimation of median and quantile models with fixed effects for ordinal outcomes can also be performed using maximum score type estimation methods that attempt to estimate $\mu _{i}$ in a large $T$ framework. However, bias correction in these settings is more challenging, which is an interesting topic for further research. 

\section*{Appendix}

This appendix shows the semiparametric median and quantile estimation problems can be formulated as mixed integer optimization problems. We start with the median. To set up the problem, note that the term $\left\vert
Y-\sum\nolimits_{j=1}^{J}j\times \mathbf{1}\left\{ \gamma _{j-1}<X^{\top
}\theta \leq \gamma _{j}\right\} \right\vert $ can be re-written as  
\begin{equation*}
\left\vert Y-J\right\vert +\sum\nolimits_{j=1}^{J-1}\left[ \left\vert
Y-j\right\vert -\left\vert Y-j-1\right\vert \right] \times \mathbf{1}\left\{
X^{\top }\theta \leq \gamma _{j}\right\}.
\end{equation*}%
This latter representation facilitates numerical evaluation of the objective function in (\ref{LAD formulation 1}) by eliminating compound inequalities in the indicator functions. The minimization problem (\ref{LAD formulation 1}) is therefore equivalent to the following minimization problem:

\begin{equation*}
\min \{\min_{(b,c)\in \Theta }S_{n}(1,b,c),\min_{(b,c)\in \Theta
}S_{n}(-1,b,c)\} ,
\end{equation*}%
where%
\begin{equation}
\label{Sn}
S_{n}(a,b,c):=\sum\nolimits_{i=1}^{n}\sum\nolimits_{j=1}^{J-1}\left[
\left\vert Y_{i}-j\right\vert -\left\vert Y_{i}-j-1\right\vert \right]
\times \mathbf{1}\left\{ aX_{1i}+\widetilde{X}_{i}^{\top }b\leq
c_{j}\right\} .
\end{equation}

Our computational algorithm solves two LAD sub-problems: 
\begin{equation}
\min_{(b,c)\in \Theta }S_{n}(a,b,c) , \label{LAD subproblem}
\end{equation}%
for $a\in \left\{ -1,1\right\} $. For each value of $a$, we can reformulate
the LAD sub-problem (\ref{LAD subproblem}) as the following mixed integer
linear programming (MILP) problem:%
\begin{eqnarray}
&&\min_{\left( b,c\right) \in \Theta
,(d_{i,1},...,d_{i,J-1})_{i=1}^{n}}\sum\nolimits_{i=1}^{n}\sum%
\nolimits_{j=1}^{J-1}\left[ \left\vert Y_{i}-j\right\vert -\left\vert
Y_{i}-j-1\right\vert \right] \times d_{i,j}  \label{LAD MIO formulation} \\
&&\text{subject to}  \notag \\
&&\left( d_{i,j}-1\right) M_{i,j}\leq c_{j}-aX_{1i}-\widetilde{X}_{i}^{\top
}b<d_{i,j}(M_{i,j}+\delta ),\text{ }\left( i,j\right) \in \{1,...,n\}\times
\{1,...,J-1\},  \label{sign constraints} \\
&&c_{j}<c_{j+1},\text{ }j\in \{1,...,J-2\},  \label{threshold monotonicity}
\\
&&d_{i,j}\leq d_{i,j+1},\text{ }\left( i,j\right) \in \{1,...,n\}\times
\{1,...,J-2\},  \label{indicator monotonicity} \\
&&d_{i,j}\in \{0,1\},\text{ }\left( i,j\right) \in \{1,...,n\}\times
\{1,...,J-1\},  \label{indicator constraints}
\end{eqnarray}%
where $\delta >0$ is a small positive scalar (e.g. we use $\delta =10^{-6}$ in
our application), and%
\begin{equation}
M_{i,j}\equiv \max\limits_{\left( b,c\right) \in \Theta }\left\vert
c_{j}-aX_{1i}-\widetilde{X}_{i}^{\top }b\right\vert ,\text{ }\left(
i,j\right) \in \{1,...n\}\times \{1,...,J-1\}.  \label{M}
\end{equation}

\bigskip 

\textsc{Proposition 3. }Solving the constrained MILP problem (\ref{LAD MIO
formulation}) is equivalent to solving the minimization problem (\ref{LAD
subproblem}).

\bigskip 

To see why Proposition 3 is true, take any $\left( b,c\right) \in \Theta $.
The sign constraints (\ref{sign constraints}) and the dichotomization
constraints (\ref{indicator constraints}) ensure that $d_{i,j}=\mathbf{1}%
\left\{ aX_{1i}+\widetilde{X}_{i}^{\top }b\leq c_{j}\right\} $ for $\left(
i,j\right) \in \{1,...n\}\times \{1,...,J-1\}$. We also enforce monotonicity
of the threshold parameters through inequality constraints (\ref{threshold
monotonicity}). Note that (\ref{sign constraints}), (\ref{threshold
monotonicity}) and (\ref{indicator constraints}) together imply (\ref%
{indicator monotonicity}), which we explicitly impose so as to further
tighten the MILP problem.

The equivalence between (\ref{LAD subproblem}) and (\ref{LAD MIO formulation}%
) enables us to employ the modern MIO solvers to exactly compute the LAD
estimator $(\widehat{\theta }_{1},\widehat{\beta },\widehat{\gamma })$. For
numerical implementation, note that the values $\left(
M_{i,1},...,M_{i,J-1}\right) _{i=1}^{n}$ in the inequality constraints (\ref%
{sign constraints}) can be computed by formulating the maximization problem
in (\ref{M}) as linear programming problems, which can be efficiently solved
by modern optimization solvers. Hence these values can be computed and
stored beforehand as the input to the MILP problem (\ref{LAD MIO formulation}).\par

The estimation of semiparametric quantile can be performed in a similar fashion with minor modifications. Specifically, it can be shown that the estimation problem (\ref{QR formulation 1}) also amounts to solving two quantile regression sub-problems taking the same form as (\ref{LAD subproblem}) but with the absolute value difference terms,  $[|Y_{i}-j|-|Y_{i}-j-1|]$, in the objective function (\ref{Sn}) being replaced by the check loss differences, $[\rho_{\alpha}(Y_{i}-j)-\rho_{\alpha}(Y_{i}-j-1)]$. Consequently, we can equivalently reformulate these sub-problems as the following MILP problems:
\begin{eqnarray}
&&\min_{\left( b,c\right) \in \Theta
,(d_{i,1},...,d_{i,J-1})_{i=1}^{n}}\sum\nolimits_{i=1}^{n}\sum%
\nolimits_{j=1}^{J-1}\left[ \rho_{\alpha}(Y_{i}-j)-\rho_{\alpha}(Y_{i}-j-1)\right] \times d_{i,j}  \label{QR MIO formulation} 
\end{eqnarray}
\noindent subject to constraints (\ref{sign constraints}), (\ref{threshold monotonicity}), (\ref{indicator monotonicity}), (\ref{indicator constraints}).

\bibliographystyle{apalike}
\bibliography{bibl}

\begin{thebibliography}{}

\bibitem[Agresti, 1999]{agresti_modelling_1999}
Agresti, A. (1999).
\newblock Modelling ordered categorical data: recent advances and future
  challenges.
\newblock {\em Statistics in Medicine}, 18(17-18):2191--2207.

\bibitem[Alesina et~al., 2004]{alesina_inequality_2004}
Alesina, A., {Di Tella}, R., and MacCulloch, R. (2004).
\newblock Inequality and happiness: are europeans and americans different?
\newblock {\em Journal of Public Economics}, 88(9):2009--2042.

\bibitem[Ananth, 1997]{ananth_regression_1997}
Ananth, C. (1997).
\newblock Regression models for ordinal responses: a review of methods and
  applications.
\newblock {\em International Journal of Epidemiology}, 26(6):1323--1333.

\bibitem[Baetschmann et~al., 2015]{baetschmann_consistent_2015}
Baetschmann, G., Staub, K.~E., and Winkelmann, R. (2015).
\newblock Consistent estimation of the fixed effects ordered logit model.
\newblock {\em Journal of the Royal Statistical Society: Series A (Statistics
  in Society)}, 178(3):685--703.

\bibitem[Bertsimas and Weismantel, 2005]{bertsimas_optimization_2005}
Bertsimas, D. and Weismantel, R. (2005).
\newblock {\em Optimization {Over} {Integers}}.
\newblock Dynamic Ideas, Belmont, Mass.

\bibitem[Bester and Hansen, 2012]{bester_penalty_2012}
Bester, C.~A. and Hansen, C. (2012).
\newblock A {Penalty} {Function} {Approach} to {Bias} {Reduction} in
  {Nonlinear} {Panel} {Models} with {Fixed} {Effects}.
\newblock {\em Journal of Business \& Economic Statistics}.
\newblock Publisher: Taylor \& Francis.

\bibitem[Blanchflower and Oswald, 2004]{blanchflower_well-being_2004}
Blanchflower, D.~G. and Oswald, A.~J. (2004).
\newblock Well-being over time in {Britain} and the {USA}.
\newblock {\em Journal of Public Economics}, 88(7-8):1359--1386.

\bibitem[Blanchflower and Oswald, 2008]{blanchflower_is_2008}
Blanchflower, D.~G. and Oswald, A.~J. (2008).
\newblock Is well-being {U}-shaped over the life cycle?
\newblock {\em Social Science \& Medicine}, 66(8):1733--1749.

\bibitem[Bloem, 2021]{bloem_how_nodate}
Bloem, J.~R. (2021).
\newblock How {Much} {Does} the {Cardinal} {Treatment} of {Ordinal} {Variables}
  {Matter}? {An} {Empirical} {Investigation}.
\newblock {\em Political Analysis}, pages 1--17.
\newblock Publisher: Cambridge University Press.

\bibitem[Bloem and Oswald, 2021]{bloem_analysis_2021}
Bloem, J.~R. and Oswald, A.~J. (2021).
\newblock The analysis of human feelings: A practical suggestion for a
  robustness test.
\newblock {\em Review of Income and Wealth}.

\bibitem[Bond and Lang, 2014]{bond_2014}
Bond, T.~N. and Lang, K. (2014).
\newblock The sad truth about happiness scales.
\newblock Working Paper 19950, National Bureau of Economic Research.

\bibitem[Bond and Lang, 2019]{bond_sad_2019}
Bond, T.~N. and Lang, K. (2019).
\newblock The {Sad} {Truth} about {Happiness} {Scales}.
\newblock {\em Journal of Political Economy}, 127(4):1629--1640.
\newblock Publisher: The University of Chicago Press.

\bibitem[Boyce et~al., 2013]{boyce_money_2013}
Boyce, C.~J., Wood, A.~M., Banks, J., Clark, A.~E., and Brown, G. D.~A. (2013).
\newblock Money, well-being, and loss aversion: does an income loss have a
  greater effect on well-being than an equivalent income gain?
\newblock {\em Psychological Science}, 24(12):2557--2562.

\bibitem[Carneiro et~al., 2003]{carneiro_estimating_2003}
Carneiro, P., Hansen, K.~T., and Heckman, J.~J. (2003).
\newblock Estimating {Distributions} of {Treatment} {Effects} with an
  {Application} to the {Returns} to {Schooling} and {Measurement} of the
  {Effects} of {Uncertainty} on {College} {Choice}.
\newblock {\em International Economic Review}, 44(2):361--422.

\bibitem[Chamberlain, 1980]{chamberlain_analysis_1980}
Chamberlain, G. (1980).
\newblock Analysis of covariance with qualitative data.
\newblock {\em Review of Economic Studies}, 47(1):225--238.

\bibitem[Chen and Lee, 2018]{chen_best_2018}
Chen, L.~Y. and Lee, S. (2018).
\newblock Best subset binary prediction.
\newblock {\em Journal of Econometrics}, 206(1):39--56.

\bibitem[Cheng et~al., 2017]{cheng_longitudinal_2017}
Cheng, T.~C., Powdthavee, N., and Oswald, A.~J. (2017).
\newblock Longitudinal {Evidence} for a {Midlife} {Nadir} in {Human}
  {Well}-being: {Results} from {Four} {Data} {Sets}.
\newblock {\em The Economic Journal}, 127(599):126--142.

\bibitem[Clark, 2003]{clark_unemployment_2003}
Clark, A.~E. (2003).
\newblock Unemployment as a {Social} {Norm}: {Psychological} {Evidence} from
  {Panel} {Data}.
\newblock {\em Journal of Labor Economics}, 21(2):289--322.

\bibitem[Clark et~al., 2008]{clark_relative_2008}
Clark, A.~E., Frijters, P., and Shields, M.~A. (2008).
\newblock Relative {Income}, {Happiness}, and {Utility}: {An} {Explanation} for
  the {Easterlin} {Paradox} and {Other} {Puzzles}.
\newblock {\em Journal of Economic Literature}, 46(1):95--144.

\bibitem[Clark and Oswald, 1994]{clark_unhappiness_1994}
Clark, A.~E. and Oswald, A.~J. (1994).
\newblock Unhappiness and {Unemployment}.
\newblock {\em The Economic Journal}, 104(424):648--659.
\newblock \_eprint:
  https://academic.oup.com/ej/article-pdf/104/424/648/27040276/ej0648.pdf.

\bibitem[Clark and Oswald, 1996]{clark_satisfaction_1996}
Clark, A.~E. and Oswald, A.~J. (1996).
\newblock Satisfaction and comparison income.
\newblock {\em Journal of Public Economics}, 61(3):359--381.

\bibitem[Conforti et~al., 2014]{conforti_integer_2014}
Conforti, M., Cornuejols, G., and Zambelli, G. (2014).
\newblock {\em Integer {Programming}}.
\newblock Graduate {Texts} in {Mathematics}. Springer International Publishing.

\bibitem[Cunha et~al., 2007]{cunha_identification_2007}
Cunha, F., Heckman, J.~J., and Navarro, S. (2007).
\newblock The {Identification} {And} {Economic} {Content} {Of} {Ordered}
  {Choice} {Models} {With} {Stochastic} {Thresholds}.
\newblock {\em International Economic Review}, 48(4):1273--1309.

\bibitem[de~Castro and Galvao, 2019]{de2019}
de~Castro, L. and Galvao, A.~F. (2019).
\newblock Dynamic quantile models of rational behavior.
\newblock {\em Econometrica}, 87(6):1893--1939.

\bibitem[De~Neve and Sachs, 2020]{de_neve_sdgs_2020}
De~Neve, J.-E. and Sachs, J.~D. (2020).
\newblock The {SDGs} and human well-being: a global analysis of synergies,
  trade-offs, and regional differences.
\newblock {\em Scientific Reports}, 10(1):15113.
\newblock Number: 1 Publisher: Nature Publishing Group.

\bibitem[Di~Tella et~al., 2001]{di_tella_preferences_2001}
Di~Tella, R., MacCulloch, R.~J., and Oswald, A.~J. (2001).
\newblock Preferences over {Inflation} and {Unemployment}: {Evidence} from
  {Surveys} of {Happiness}.
\newblock {\em American Economic Review}, 91(1):335--341.

\bibitem[Easterlin, 1974]{easterlin_does_1974}
Easterlin, R.~A. (1974).
\newblock Does {Economic} {Growth} {Improve} the {Human} {Lot}? {Some}
  {Empirical} {Evidence}.
\newblock In David, P.~A. and Reder, M.~W., editors, {\em Nations and
  {Households} in {Economic} {Growth}}, pages 89--125. Academic Press.

\bibitem[Easterlin, 2003]{easterlin_explaining_2003}
Easterlin, R.~A. (2003).
\newblock Explaining happiness.
\newblock {\em Proceedings of the National Academy of Sciences},
  100(19):11176--11183.

\bibitem[{Ferrer-i-Carbonell}, 2005]{ferrer-i-carbonell_income_2005}
{Ferrer-i-Carbonell}, A. (2005).
\newblock Income and well-being: an empirical analysis of the comparison income
  effect.
\newblock {\em Journal of Public Economics}, 89(5-6):997--1019.

\bibitem[{Ferrer-i-Carbonell} and Frijters, 2004]{ferrericarbonell_how_2004}
{Ferrer-i-Carbonell}, A. and Frijters, P. (2004).
\newblock How {Important} is {Methodology} for the estimates of the
  determinants of {Happiness}?
\newblock {\em The Economic Journal}, 114(497):641--659.

\bibitem[Florios and Skouras, 2008]{florios_exact_2008}
Florios, K. and Skouras, S. (2008).
\newblock Exact computation of max weighted score estimators.
\newblock {\em Journal of Econometrics}, 146(1):86--91.

\bibitem[Frey and Stutzer, 2000]{frey_happiness_2000}
Frey, B.~S. and Stutzer, A. (2000).
\newblock Happiness, economy and institutions.
\newblock {\em The Economic Journal}, 110(466):918--938.

\bibitem[Gebers, 1998]{gebers_exploratory_1998}
Gebers, M. (1998).
\newblock Exploratory {Multivariable} {Analyses} of {California} {Driver}
  {Record} {Accident} {Rates}.
\newblock {\em Transportation Research Record: Journal of the Transportation
  Research Board}, 1635:72--80.

\bibitem[Greene and Hensher, 2010]{greene_hensher_2010}
Greene, W.~H. and Hensher, D.~A. (2010).
\newblock {\em Modeling Ordered Choices: A Primer}.
\newblock Cambridge University Press.

\bibitem[Gruber and Mullainathan, 2006]{gruber_cigarette_2006}
Gruber, J. and Mullainathan, S. (2006).
\newblock Do {Cigarette} {Taxes} {Make} {Smokers} {Happier}?
\newblock In Ng, Y.~K. and Ho, L.~S., editors, {\em Happiness and {Public}
  {Policy}: {Theory}, {Case} {Studies} and {Implications}}, pages 109--146.
  Palgrave Macmillan UK, London.

\bibitem[Hahn and Newey, 2004]{hahn_jackknife_2004}
Hahn, J. and Newey, W. (2004).
\newblock Jackknife and {Analytical} {Bias} {Reduction} for {Nonlinear} {Panel}
  {Models}.
\newblock {\em Econometrica}, 72(4):1295--1319.
\newblock \_eprint:
  https://onlinelibrary.wiley.com/doi/pdf/10.1111/j.1468-0262.2004.00533.x.

\bibitem[Honore and Lewbel, 2002]{honore_semiparametric_2002}
Honore, B.~E. and Lewbel, A. (2002).
\newblock Semiparametric {Binary} {Choice} {Panel} {Data} {Models} {Without}
  {Strictly} {Exogeneous} {Regressors}.
\newblock {\em Econometrica}, 70(5):2053--2063.

\bibitem[Horowitz, 1992]{horowitz_smoothed_1992}
Horowitz, J.~L. (1992).
\newblock A {Smoothed} {Maximum} {Score} {Estimator} for the {Binary}
  {Response} {Model}.
\newblock {\em Econometrica}, 60(3):505--531.

\bibitem[Horowitz, 2009]{horowitz_semiparametric_2009}
Horowitz, J.~L. (2009).
\newblock {\em Semiparametric and {Nonparametric} {Methods} in {Econometrics}}.
\newblock Springer {Series} in {Statistics}. Springer-Verlag, New York.

\bibitem[Kahneman and Krueger, 2006]{kahneman_developments_2006}
Kahneman, D. and Krueger, A.~B. (2006).
\newblock Developments in the {Measurement} of {Subjective} {Well}-{Being}.
\newblock {\em Journal of Economic Perspectives}, 20(1):3--24.

\bibitem[Kahneman et~al., 2004]{kahneman_toward_2004}
Kahneman, D., Krueger, A.~B., Schkade, D., Schwarz, N., and Stone, A. (2004).
\newblock Toward {National} {Well}-{Being} {Accounts}.
\newblock {\em American Economic Review}, 94(2):429--434.

\bibitem[Kaiser and Vendrik, 2020]{kaiser_how_nodate}
Kaiser, C. and Vendrik, M. C.~M. (2020).
\newblock How threatening are transformations of happiness scales to subjective
  wellbeing research?

\bibitem[Kaplan and Zhuo, 2020]{KaplanZhuo2021b}
Kaplan, D.~M. and Zhuo, L. (2020).
\newblock Comparing latent inequality with ordinal data.

\bibitem[Kim and Pollard, 1990]{kim_cube_1990}
Kim, J. and Pollard, D. (1990).
\newblock Cube {Root} {Asymptotics}.
\newblock {\em The Annals of Statistics}, 18(1):191--219.

\bibitem[King et~al., 2004]{KinMurSal04}
King, G., Murray, C.~J., Salomon, J.~A., and Tandon, A. (2004).
\newblock Enhancing the validity and cross-cultural comparability of
  measurement in survey research.
\newblock {\em American Political Science Review}, 98:191{\textendash}207.

\bibitem[Koenker and Bassett, 1978]{koenker_simple_1978}
Koenker, R. and Bassett, G. (1978).
\newblock Regression quantiles.
\newblock {\em Econometrica}, 46(1):33--50.

\bibitem[Kohler et~al., 2005]{kohler_population_2005}
Kohler, H.-P., Behrman, J.~R., and Skytthe, A. (2005).
\newblock Partner + children = happiness? the effects of partnerships and
  fertility on well-being.
\newblock {\em Population and Development Review}, 31(3):407--445.

\bibitem[Kotlyarova and Zinde-Walsh, 2009]{kotlyarova_robust_2009}
Kotlyarova, Y. and Zinde-Walsh, V. (2009).
\newblock Robust {Estimation} in {Binary} {Choice} {Models}.
\newblock {\em Communications in Statistics - Theory and Methods},
  39(2):266--279.

\bibitem[Lee, 1992]{lee_median_1992}
Lee, M.~J. (1992).
\newblock Median regression for ordered discrete response.
\newblock {\em Journal of Econometrics}, 51(1-2):59--77.

\bibitem[Lewbel, 1997]{lewbel_constructing_1997}
Lewbel, A. (1997).
\newblock Constructing {Instruments} for {Regressions} with {Measurement}
  {Error} when no {Additional} {Data} are {Available}, with an {Application} to
  {Patents} and {R}\&{D}.
\newblock {\em Econometrica}, 65(5):1201--1214.

\bibitem[Lewbel, 2000]{lewbel_semiparametric_2000}
Lewbel, A. (2000).
\newblock Semiparametric qualitative response model estimation with unknown
  heteroscedasticity or instrumental variables.
\newblock {\em Journal of Econometrics}, 97(1):145--177.

\bibitem[Lewbel and Schennach, 2007]{lewbel_simple_2007}
Lewbel, A. and Schennach, S.~M. (2007).
\newblock A simple ordered data estimator for inverse density weighted
  functions.
\newblock {\em Journal of Econometrics}, 136:189--211.

\bibitem[Ludwig et~al., 2012]{ludwig_neighborhood_2012}
Ludwig, J., Duncan, G.~J., Gennetian, L.~A., Katz, L.~F., Kessler, R.~C.,
  Kling, J.~R., and Sanbonmatsu, L. (2012).
\newblock Neighborhood effects on the long-term well-being of low-income
  adults.
\newblock {\em Science}, 337(6101):1505--1510.

\bibitem[Luttmer, 2005]{luttmer_neighbors_2005}
Luttmer, E. (2005).
\newblock Neighbors as negatives: Relative earnings and well-being.
\newblock {\em The Quarterly Journal of Economics}, 120(3):963--1002.

\bibitem[Major, 2012]{major_timing_2012}
Major, S. (2012).
\newblock Timing {Is} {Everything}: {Economic} {Sanctions}, {Regime} {Type},
  and {Domestic} {Instability}.
\newblock {\em International Interactions}, 38(1):79--110.

\bibitem[Manski, 1975]{manski_maximum_1975}
Manski, C.~F. (1975).
\newblock Maximum score estimation of the stochastic utility model of choice.
\newblock {\em Journal of Econometrics}, 3(3):205--228.

\bibitem[Manski, 1985]{manski_semiparametric_1985}
Manski, C.~F. (1985).
\newblock Semiparametric analysis of discrete response: {Asymptotic} properties
  of the maximum score estimator.
\newblock {\em Journal of Econometrics}, 27(3):313--333.

\bibitem[Manski, 1988a]{manski_identification_1988}
Manski, C.~F. (1988a).
\newblock Identification of {Binary} {Response} {Models}.
\newblock {\em Journal of the American Statistical Association},
  83(403):729--738.

\bibitem[Manski, 1988b]{manski1988}
Manski, C.~F. (1988b).
\newblock Ordinal utility models of decision making under uncertainty.
\newblock {\em Theory and Decision}, 25(1):79--104.

\bibitem[Manski and Thompson, 1986]{manski_operational_1986}
Manski, C.~F. and Thompson, T.~S. (1986).
\newblock Operational characteristics of maximum score estimation.
\newblock {\em Journal of Econometrics}, 32(1):85--108.

\bibitem[McCullagh, 1980]{mccullagh_regression_1980}
McCullagh, P. (1980).
\newblock Regression {Models} for {Ordinal} {Data}.
\newblock {\em Journal of the Royal Statistical Society. Series B
  (Methodological)}, 42(2):109--142.

\bibitem[McKelvey and Zavoina, 1975]{mckelvey_statistical_1975}
McKelvey, R.~D. and Zavoina, W. (1975).
\newblock A statistical model for the analysis of ordinal level dependent
  variables.
\newblock {\em The Journal of Mathematical Sociology}, 4(1):103--120.

\bibitem[Oparina and Srisuma, 2022]{oparina_analyzing_2021}
Oparina, E. and Srisuma, S. (2022).
\newblock Analyzing subjective well-being data with misclassification.
\newblock {\em Journal of Business \& Economic Statistics}, 40(2):730--743.

\bibitem[Pinkse, 1993]{pinkse_computation_1993}
Pinkse, C. A.~P. (1993).
\newblock On the computation of semiparametric estimates in limited dependent
  variable models.
\newblock {\em Journal of Econometrics}, 58(1):185--205.

\bibitem[Powdthavee, 2007]{powdthavee_are_2007}
Powdthavee, N. (2007).
\newblock Are there {Geographical} {Variations} in the {Psychological} {Cost}
  of {Unemployment} in {South} {Africa}?
\newblock {\em Social Indicators Research: An International and
  Interdisciplinary Journal for Quality-of-Life Measurement}, 80(3):629--652.

\bibitem[Rostek, 2010]{rostek2010}
Rostek, M. (2010).
\newblock Quantile maximization in decision theory.
\newblock {\em The Review of Economic Studies}, 77(1):339--371.

\bibitem[{Schr\"{o}der} and Yitzhaki, 2017]{schroder_revisiting_2017}
{Schr\"{o}der}, C. and Yitzhaki, S. (2017).
\newblock Revisiting the evidence for cardinal treatment of ordinal variables.
\newblock {\em European Economic Review}, 92:337--358.

\bibitem[Sechel, 2021]{sechel_share_2021}
Sechel, C. (2021).
\newblock The share of satisfied individuals: A headcount measure of aggregate
  subjective well-being.
\newblock {\em Journal of Economic Behavior \& Organization}, 186:373--394.

\bibitem[Seo and Otsu, 2018]{seo_local_2018}
Seo, M.~H. and Otsu, T. (2018).
\newblock Local {M}-estimation with discontinuous criterion for dependent and
  limited observations.
\newblock {\em The Annals of Statistics}, 46(1):344--369.

\bibitem[Skouras, 2003]{skouras_algorithm_2003}
Skouras, S. (2003).
\newblock An algorithm for computing estimators that optimize step functions.
\newblock {\em Computational Statistics \& Data Analysis}, 42(3):349--361.

\bibitem[Stevenson and Wolfers, 2009]{stevenson_paradox_2009}
Stevenson, B. and Wolfers, J. (2009).
\newblock The {Paradox} of {Declining} {Female} {Happiness}.
\newblock {\em American Economic Journal: Economic Policy}, 1(2):190--225.

\bibitem[Stevenson and Wolfers, 2013]{stevenson_subjective_2013}
Stevenson, B. and Wolfers, J. (2013).
\newblock Subjective {Well}-{Being} and {Income}: {Is} {There} {Any} {Evidence}
  of {Satiation}?
\newblock {\em American Economic Review}, 103(3):598--604.

\bibitem[Stiglitz et~al., 2009]{stiglitz_measurement_2009}
Stiglitz, J., Sen, A.~K., and Fitoussi, J.~P. (2009).
\newblock The measurement of economic performance and social progress
  revisited: {Reflections} and {Overview}.
\newblock Technical Report 2009-33, Sciences Po.

\bibitem[Stutzer and Frey, 2013]{stutzer_recent_2013}
Stutzer, A. and Frey, B.~S. (2013).
\newblock Recent {Developments} in the {Economics} of {Happiness}: {A}
  {Selective} {Overview}.
\newblock In {\em Recent {Developments} in the {Economics} of {Happiness}}.
  Cheltenham.

\bibitem[Williams, 2010]{williams_fitting_2010}
Williams, R. (2010).
\newblock Fitting heterogeneous choice models with oglm.
\newblock {\em The Stata Journal}, 10(4):540--567.

\end{thebibliography}

\end{document}